**Noisy probing dose facilitated dose prediction for pencil beam scanning proton therapy: physics enhances generalizability**

Lian Zhang, PhD[1*], Jason M. Holmes, PhD[1*], Zhengliang Liu, MS[2], Hongying Feng, PhD[1], Terence T. Sio, MD, MS[1], Carlos E. Vargas, MD[1], Sameer R. Keole, MD[1], Kristin Stützer, PhD[3,4], Sheng Li, PhD[5], Tianming Liu, PhD[2], Jiajian Shen, PhD[1], William W. Wong, MD[1], Sujay A. Vora, MD[1], Wei Liu, PhD[1]

[1]Department of Radiation Oncology, Mayo Clinic, Phoenix, AZ 85054, USA

[2]School of Computing, University of Georgia, Athens, GA 30602, USA

[3]Oncoray – National Center for Radiation Research in Oncology, Faculty of Medicin and University Hospital Carl Gustav Carus, Technische Universität Dresden, Helmholtz-Zentrum Dresden - Rossendorf, Dresden, Germany

[4]Helmholtz-Zentrum Dresden - Rossendorf, Institute of Radiooncology – OncoRay, Dresden, Germany

[5]School of Data Science, University of Virginia, Charlottesville, VA 22903, USA

[*]Co-first authors who contribute to this paper equally

Corresponding author: Wei Liu, PhD, Professor of Radiation Oncology, Department of Radiation Oncology, Mayo Clinic Arizona, 5777 E. Mayo Boulevard, Phoenix, AZ 85054; e-mail: Liu.Wei@mayo.edu.

**Acknowledgments**

This research was supported by the National Cancer Institute (NCI) Career Developmental Award K25CA168984, the Arizona Biomedical Research Commission Investigator Award, the Lawrence W. and Marilyn W. Matteson Fund for Cancer Research, and the Kemper Marley Foundation.

**Conflicts of Interest Notification**

Terence T. Sio provides strategic and scientific recommendations as a member of the Advisory Board and speaker for Novocure, Inc., Catalyst Pharmaceuticals, Inc., and Galera Pharmaceutics, which are not in any way associated with the content presented in this manuscript.

**Ethical considerations**

This research was approved by the Mayo Clinic Arizona institutional review board (IRB, 13-005709). The informed consent was waived by IRB protocol. Only CT image and dose-volume data were used in this study. All patient-related health information was removed prior to the analysis and publication of the study.

Data Availability Statement

The data are available from the corresponding author upon reasonable request.

**Abstract:**

Purpose: Prior AI-based dose prediction studies in photon and proton therapy often neglect underlying physics, limiting their generalizability to handle outlier clinical cases, especially for pencil beam scanning proton therapy (PBSPT). Our aim is to design a physics-aware and generalizable AI-based PBSPT dose prediction method that has the underlying physics considered to achieve high generalizability to properly handle the outlier clinical cases.

Methods and Materials: This study analyzed PBSPT plans of 103 prostate (93 for training and 10 for testing) and 78 lung cancer patients (68 for training and 10 for testing) from our institution, with each case comprising CT images, structure sets, and plan doses from our Monte-Carlo dose engine (serving as the ground truth). Three methods were evaluated in the ablation study: the ROI-based method, the beam mask and sliding window method, and the noisy probing dose method. Twelve cases with uncommon beam angles or prescription doses tested the methods' generalizability to rare treatment planning scenarios. Performance evaluation used dose-volume histogram (DVH) indices, 3D Gamma passing rates (criteria: 3%/2mm/10%), and dice coefficients for dose agreement. Prediction times for test cases gauged model efficiency.

Results: The noisy probing dose method showed improved agreement of DVH indices, 3D Gamma passing rates, and dice coefficients compared to the conventional methods for the testing cases. The noisy probing dose method showed better generalizability in the 6 outlier cases than the ROI-based and beam mask-based methods with 3D Gamma passing rates (for prostate cancer, targets:

89.32%±1.45% vs. 93.48%±1.51% vs. 96.79%±0.83%, OARs: 85.87%±1.73% vs. 91.15%±1.13% vs. 94.29%±1.01%). The dose predictions for all testing cases were completed within 0.3 seconds.

Conclusions: We've devised a novel noisy probing dose method for PBSPT dose prediction in prostate and lung cancer patients. With more physics included, it enhances the generalizability of dose prediction in handling outlier clinical cases.

## 1. Introduction

The state-of-the-art beam delivery technique in proton therapy today is pencil beam scanning proton therapy (PBSPT). PBSPT is highly adaptable compared to prior proton therapy delivery techniques while reducing dose exposure to normal tissues as compared to modern X-ray-based radiation therapy[1-8]. Despite these benefits, PBSPT is more sensitive to range and setup uncertainties in clinical settings compared to X-ray-based modalities, therefore requiring highly accurate dose calculation techniques[9-14]. Additionally, with emerging techniques such as online-adaptive radiotherapy (ART), FLASH, grid therapy, RBE-based dose optimization, etc., and especially considering robust optimization, there is a constant need to increase the speed of treatment planning for PBSPT while not sacrificing accuracy[15-24]. In general, faster treatment planning enables new capabilities to be explored that were previously considered infeasible[25-28].

Artificial intelligence (AI) may hold the key to further increasing treatment planning speed, especially for optimal planning dose generation[29]. AI-based models hold a unique advantage over conventional optimal planning dose generation methods: they are fast and typically finish highly complex tasks in less than a second[30-32]. AI-based methods have been recently studied for dose prediction in photon-based modalities. Preliminary research has incorporated AI-based methods to facilitate treatment planning by estimating dose distributions for 'optimal' schemes in specific contexts, drawn from past treatment plan dose distribution data[29]. The majority of these AI-based methods utilize organ and target masks and CT images as input to directly predict dose distributions[33-37]. The effectiveness of these advanced AI-based methods in predicting voxel-wise dose heavily depends on the data used in the model training. Model developers must judiciously select patients who have consistent beam configurations and/or dose prescriptions[29]. Such careful training data selection, therefore, is crucial for developing a precise dose prediction model.

However, this limits the model's generalizability to adapt to more diverse beam configurations and/or dose prescriptions, a key problem in dose prediction for PBSPT where the number of beams and their angles can significantly differ from one patient to another especially for some uncommon treatment planning cases, or between different institutions due to institution specific treatment planning protocols. Hence, the practicality of implementing automatic treatment planning based on these models in clinical scenarios seems prohibitable, given the necessity to create distinct models for every single beam layout.

To enhance the overall accuracy, generalizability, and the interpretability of the dose prediction models, the input data may need to be expanded to include patient-specific physics information, which is the subject of this work. To date, there have been primarily three different approaches for AI-based dose prediction (optimal planning dose generation) in radiation therapy: (1) to use the past treatment plan data only (such as the dose distribution, CT images, region of interest (ROI] contours) without the consideration of the evolving physics, culminating in the development of a purely AI-based dose prediction engine[33-36,38,39], (2) to utilize existing conventional (analytical or Monte Carlo simulation) dose calculation techniques and optimization processes to generate an inaccurate or noisy dose distribution followed by an AI-based process to rectify or de-noise the dose distribution respectively[40-43], and (3) to use beam configuration information, such as the beam angle in addition to the conventional data used in dose prediction[44-47]. Approach 1 could lead to high prediction accuracy if a prohibited number of prior treatment plan data is available to take data diversity and completeness into account. However, the model lacks physics interpretability and raises concerns about its physical reasonability of its output[33-36,38,39]. For approach 2, although the inaccurate or noisy dose distribution obtained based on the conventional processes exhibits some physics interpretability, it requires a complicated and time-consuming process to calculate and optimize the dose distribution using conventional methods, which undermines the main benefit of AI-based dose prediction, i.e., speed[40-43]. Approach 3 partially mitigates the challenges existing for approach 1. However, incorporating beam configuration data, such as the beam angle, is challenging since its inclusion necessitates diverse and heterogeneous training data spanning the parameter space of interest. This is often not possible by the data from past patients of only one single institution and data augmentation is likely to be required[44-47]. The resulting model still lacks physics interpretability as well.

While numerous studies have investigated the application of AI-based methods in dose prediction for photon therapy, existing literature on AI-based dose prediction for PBSPT remains limited[38,39,46,48]. This is due to the challenging inherent sensitivity of proton dose distributions to anatomical heterogeneities. It is therefore expected that dose distributions for photons could be accurately predicted only based on geometric considerations, whereas the dose prediction for protons is much harder since the PBSPT dose distributions vary significantly between patients, even for the same disease site, due to the different physics characteristics of protons compared to photons.

Our objective is to establish a comprehensive dose prediction model capable of adapting to diverse beam configurations. Such a versatile model could fully exploit the AI' potential for dose prediction, bringing us a step closer to the clinical implementation of automated treatment planning based on these methodologies. In a recent study on PBSPT dose prediction, an AI-based dose prediction model using a largely traditional approach was developed and was augmented with a meticulously selected beam mask and sliding window technique during model training[46]. In this study, we explore a novel approach that incorporates physics information into the input, called as the "noisy probing dose". It is derivable from the energy layer and spot spacing alone and imparts valuable physical information to the model, circumventing the need for complex dose calculation or optimization procedures. Compared to previous studies, our noisy probing dose method exhibits superior dose prediction accuracy and generalizability in PBSPT across all tested cases, including the outlier clinical cases, which demonstrates its enhanced applicability for routine clinical use.

## 2. Material and Methods

The main concept of this study is to probe the dose before predicting dose: doses from a plan with spots of the same monitor units (MUs) and in each energy layer are calculated by a low statistics proton Monte Carlo simulation. A conventional complicated optimization process of spot weights is not needed. We call this method "noisy dose probing". The resultant noisy probing dose distribution is then used as input to the dose prediction model, providing additional physics information as well as constraining (or bounding) the model. In this way, the model is forced to predict dose in-line with the diverse beam settings that are specified ahead of time (by way of simulating them in this trivial manner). The probing dose embeds the beam angles as well as proton-tissue interaction physics information into an AI readable image (as opposed to text or some other ill-suited representation), allowing for a natural incorporation of the input into the existing 3D U-net framework. The difference from conventional treatment planning is shown in Figure 1(a), where the dose influence matrix (Dij) calculation and subsequent inverse optimization are replaced by a probing dose calculation and AI-based optimal dose prediction. Dij represents the contribution of $j$th spot with unit intensity at $i$th voxel in the region of interests, which can be considered as spot-by-spot dose distributions in proton therapy. Dose to each voxel is then calculated by adding up the contributions from all spots with proper intensities.

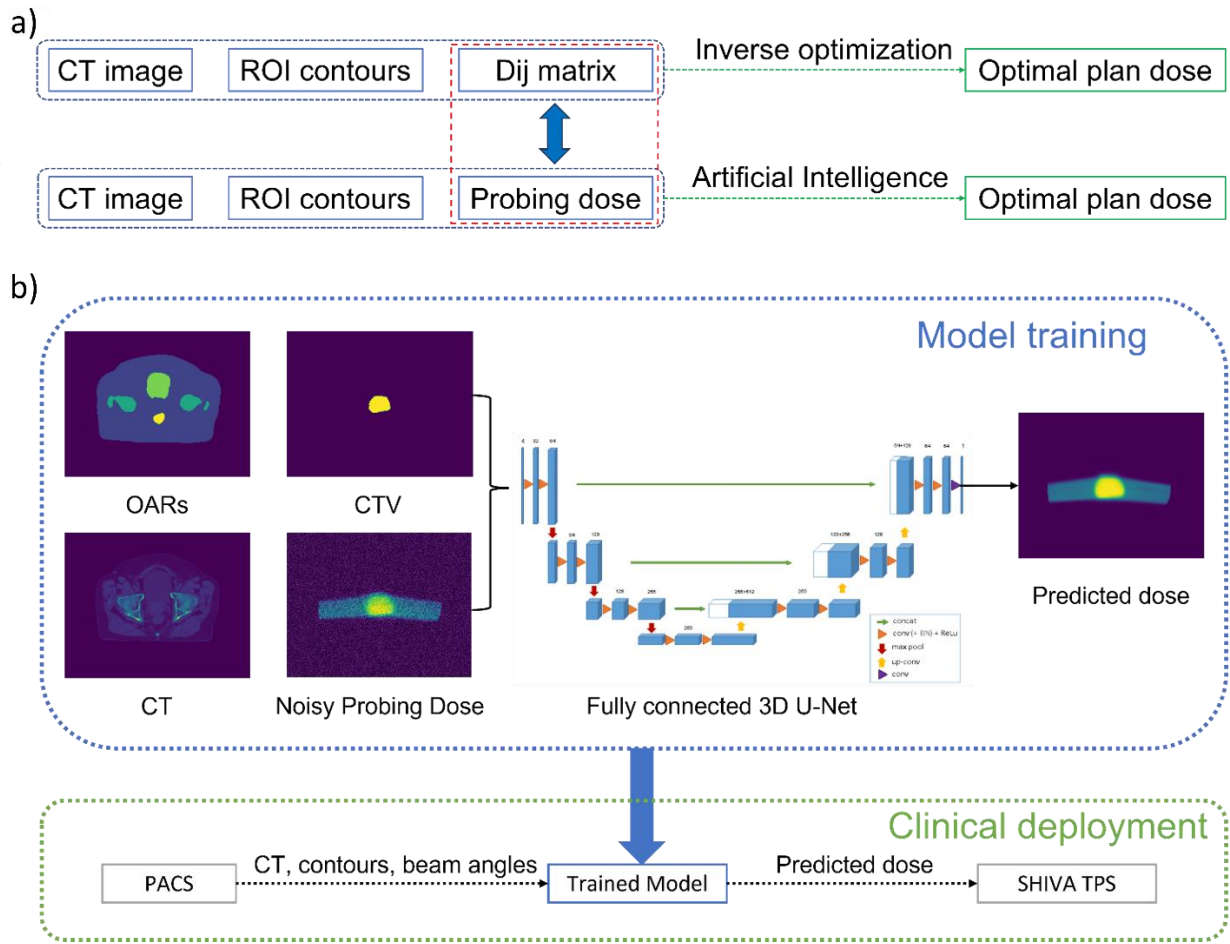

Figure 1. Workflow of proposed plan dose generation. a) AI workflow with probing dose (bottom) compared to the traditional workflow with influence matrix and inverse optimization (top). b) Diagram of noisy probing dose facilitated AI-based dose prediction for pencil beam scanning proton therapy.

2.1 Data Collection

For this study, we selected 103 prostate cancer patients and 78 lung cancer patients who were previously treated with PBSPT at our facility using our in-house patient search engine. For each disease sites, we randomly chose 10 patients as the testing dataset while the rest were allocated for model training. The target and organ at risk (OAR) delineations were examined and approved by experienced radiation oncologists. The prescribed dose was 70.0 Gy[RBE] in 28 fractions for prostate cancer, while for lung cancer it ranged from 50.0 Gy[RBE] to 60.0 Gy[RBE], administered over 25 to 30 fractions. Individualized treatment plans were developed for each patient, taking into account their unique anatomy, using different beam angles and configurations. The in-house treatment planning system (TPS), Shiva[8,12,16,22,23,26], employing a Virtual Particle Monte Carlo (VPMC) dose engine[49], was used to generate the planning dose distributions for all the patients involved in this study. Dose-volume constraints for both prostate and lung cases adhered to the clinical protocols of our institution.

## 2.2 Data Preprocessing

First, the CT scans, structure sets, and dose DICOM files computed by VPMC (considered to be the ground truth dose distributions) for each patient were extracted and converted into 3D matrices. These 3D matrices were first resampled to a 2.5 mm grid from their original resolution, and subsequently rigidly aligned to a reference case chosen for the corresponding site to ensure data consistency. We normalized the CT HU number and the dose matrices to have a mean value of 0 and a variance of 1. A bounding box of dimensions $350 \times 450 \times 550$, centered on the target, was used to crop the images, ensuring that all regions potentially impacting the dose distribution were included in the model training for all data. Zero padding was used in instances where the processed matrices' dimension was smaller than the cropping box.

Next, we generated a 3D binary bitmask for each ROI, designating a value of 1 for voxels within the contour and 0 for those outside. To differentiate the ROIs, we assigned each a specific integer number. For prostate cases, ROIs comprised CTV, bladder, spaceOAR, left/right femoral head, penile bulb, and rectum. For lung cases, they included CTV, spinal cord, spinal cord prv, esophagus, heart, and total lung. Therefore, after preprocessing, each patient's aligned and cropped data was represented by CT matrices, contour mask matrices, and dose matrices.

## 2.3 Generation of Noisy Probing Dose

Given the substantial impact of beam configurations on dose distributions in PBSPT, we proposed the noisy dose probing method to enhance the physics awareness of the AI model. First, a so-called spot-placement target volume (STV) was generated, as is typical for proton therapy treatment planning at our clinic. The STV is formed by a uniform expansion of planning target volume (PTV) for initial spot arrangement to allow for at least one spot outside the PTV. Next, spots were placed throughout the STV, layer by layer, with a constant spot spacing. With each spot weight set to 1, the probing dose is calculated using a low number of protons with VPMC. Since multiple scenarios are not needed and a low number of protons are used (1/1000 of the number of protons commonly used for conventional influence matrix calculation), dose probing can be thus performed much faster and with much less memory as compared to calculating the dose influence matrix in conventional treatment planning. A conventional optimization process to optimize spot weights is not needed either.

## 2.4 Framework of the Model and Model Training

Our chosen model framework was a fully integrated 3D U-Net, capable of handling multi-channels of 3D matrix input. These 3D input matrices were composed of the pre-processed CT matrices, contour bitmask matrices, and noisy probing dose matrices. Their combination varied based on the applied training strategy. The model's objective was to generate a 3D dose matrix output aligned with the ground-truth dose matrix. In terms of clinical trials and implementation, this trained model can be integrated into our in-house TPS, Shiva[10,11], to evaluate the predicted planning dose and facilitate potential replanning, as illustrated in Figure 1(b).

For the model training, the Smooth L1 loss function served as the objective, quantifying the discrepancy between the predicted and ground truth dose distributions. Mathematically, it is defined as:

$$L(d, \hat{d}) = \frac{1}{N} \sum_{i=1}^{N} \text{smooth}_{L_1}(d_i - \hat{d}_i) \tag{1}$$

Wherein, the Smooth L1 is given by:

$$\text{Smooth}_{L_1}(x) = f(x) = \begin{cases} 0.5x^2, & if\ |x| < 1 \\ |x| - 0.5, & \text{otherwise} \end{cases} \tag{2}$$

Here, $L$ denotes the total loss, $d_i$ is the normalized ground truth dose at the $i_{th}$ voxel, and $\hat{d}_i$ represents the predicted dose at the same voxel. Optimization was performed using the Adam optimizer, with the Reduce-LR-On-Plateau learning rate schedule in the PyTorch framework. To maximize the retention of global contextual information, full 3D pre-processed matrices as described in the data preprocessing subsection were utilized as input channels instead of patch-based method dissecting the full matrix into smaller, localized patches for the model training. Data augmentation techniques with random rotations and translations were incorporated to avoid overfitting. The models were trained for 200 epochs on a GPU workstation consisting of four NVIDIA Tesla A100 GPUs, each boasting 80 GB of onboard RAM.

2.5 Ablation Study

To assess the impact of each component of the proposed strategies, three experiments were designed. Experiment 1 utilized CT images and contour bitmasks as input channels for model training as used in the conventional ROI-based method. Experiment 2 adds the beam mask and sliding window technique to the CT images and contour bitmasks from Experiment 1. The 3D beam masks were created using raytracing, and the sliding window approach, which randomly masks out a cube of 3x3x3 voxels for each batch during training, prompts the model to focus on improving fine details in subsequent batches, ultimately enhancing the overall dose prediction precision across the dataset.[32,46]. Experiment 3 added the noisy probing dose technique to the model training, besides the methods used in Experiment 2 defined as the noisy probing dose method. A set of 10 prostate cancer and 10 lung cancer cases were used to evaluate the performance of each experiment. The ablation study allows for measuring the improvement in dose prediction versus the method used.

In assessing the accuracy of the dose prediction, three distinct evaluation metrics were employed. Firstly, DVH indices were used which include D2 and D98 for the CTV, mean dose (Dmean) for bladder, spaceOAR, femoral heads, penile bulb, rectum, esophagus, heart, and total lung, and maximum dose (Dmax) for spinal cord, spinal cord prv. Next, we employed the global 3D Gamma passing rate with a criterion of 3%/2mm/10% to demonstrate the 3D spatial dose distribution consistency between the predicted dose and the ground-truth dose in targets, ROI, and BODY (outside all ROIs), respectively. Wilcoxon signed rank tests were employed to evaluate the statistical differences in DVH indices and 3D Gamma passing rates among the three experiments: Experiment 1 (ROI method) vs. Experiment 2 (Beam mask and sliding window method), Experiment 2 vs. Experiment 3 (Noisy probing dose method), and Experiment 1 vs. Experiment 3. Lastly, we evaluated the 3D spatial dose distribution consistency through the Dice coefficients of the volumes enclosed by the iso-dose lines (with dose values ranging from 10% to 90% of the prescription dose with an increment of 10%) between the predicted and the ground truth dose distributions. The Dice coefficient indicates the volumetric similarity between the two volumes.

## 2.6 Uncommon Cases Test

In order to further test the generalizability of the model, we selectively identified six uncommon prostate and lung clinical cases: two plans with different dose prescriptions not included in the model training data, two plans with different beam numbers and angles not included in the model training data, and two plans with different prescriptions and beam configurations not included in the training data. To test the performance of the three models corresponding to three experiments in the ablation study, we generated the predicted dose distributions using the three models from Experiment 1, 2, and 3 and calculated the Gamma passing rates (3%/2mm/10%) relative to the ground truth dose distributions in the target, OARs, and BODY regions for each of these six patients. Wilcoxon signed rank tests were employed to evaluate the statistical differences in 3D Gamma passing rates among the three experiments.

Table 1. Configuration of model training for the conventional ROI model, the beam mask and sliding window model, and the noisy probing dose model.

| | ROI model | Beam mask model | Noisy probing dose model |
|---|---|---|---|
| Epoch | 200 | 200 | 200 |
| Batch size | 2 | 2 | 2 |
| Training time | 7.3 h | 7.5 h | 7.8 h |
| Predicting time | 0.259 s | 0.263 s | 0.271 s |

## 3. Results

### 3.1 Dose Distribution Comparison

Figure 2(a)/(c) provides a graphical representation of the predicted dose distributions derived from each of the three experiments. Additionally, it depicts the ground truth dose in the axial plane for a typical prostate cancer case (Fig. 2(a)) and in the sagittal plane for a typical lung cancer case (Fig. 2(c)). The figure further illustrates the differences in dose distribution within the respective planes between the predicted and the ground truth dose for the example patients.

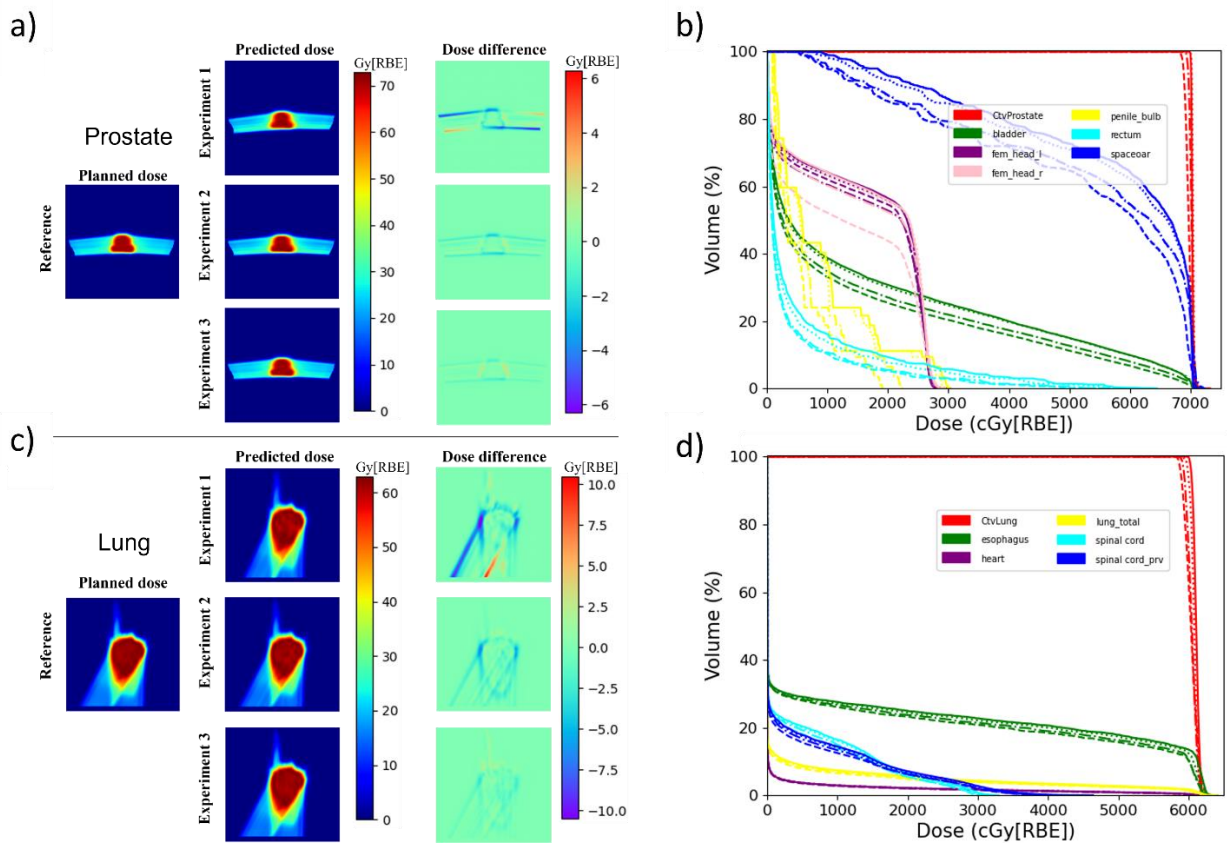

Figure 2. Comparisons of the predicted dose distributions as derived from three distinct experiments juxtaposed against the ground truth doses. a) An example prostate case in the axial plane. b) The DVH of an example prostate case. c) An example lung case in the sagittal plane. d)

The DVH of an example lung case. *Solid line*: ground-truth dose; *dash line*: predicted dose from Experiment 1, *dash dot line*: predicted dose from Experiment 2; *dot line*: predicted dose from Experiment 3.

### 3.2 DVH Comparison

Figure 2(b)/(d) displays the DVHs of the predicted doses from the three distinct experiments with the ground truth doses for both a representative prostate (Fig. 2(b)) and lung (Fig. 2(d)) cancer patient. The figures indicate an improved accuracy in dose prediction when the beam mask and sliding window model was employed relative to the ROI model, with further enhancement observed when the noisy probing dose model was used, as compared to the beam mask and sliding window method.

Figure 3(a) highlights the absolute variations in the DVH indices pertaining to the targets and OARs between the ground truth (reference) dose and the predicted doses from the three separate experiments for prostate and lung test cases. Specifically, the absolute deviation (mean ± standard deviation) for prostate CTV D98 was found to be 0.53±0.22 Gy[RBE], 0.41±0.28 Gy[RBE], and 0.29±0.06 Gy[RBE] for Experiments 1, 2, and 3, respectively. For lung CTV D98, the absolute deviations were observed to be 0.74±0.18 Gy[RBE], 0.54±0.19 Gy[RBE], and 0.34±0.12 Gy[RBE] for the aforementioned experiments in sequence. Furthermore, both prostate and lung CTV D2 from Experiment 1 exhibited absolute deviations of 0.81±0.25 Gy[RBE] and 0.94±0.44 Gy[RBE], respectively, which decreased by an average of 0.21 Gy[RBE] and 0.32 Gy[RBE] in Experiment 2, and saw a further average decrement of 0.13 Gy[RBE] in Experiment 3.

Regarding the DVH indices for OARs, the absolute variations in Dmean and Dmax between the ground truth (reference) and the predicted dose distributions originating from the three experiments were all within the clinically acceptable limits for both prostate and lung cancer patients. For the majority of OARs, the mean absolute Dmean deviation in Experiment 2 was approximately 0.15 Gy[RBE] lower than that in Experiment 1. Experiment 3 further reduced the Dmean absolute deviation on average by 0.1 Gy[RBE] compared to Experiment 2. For the lung cancer patients, the Dmax of the spinal cord and spinal cord prv exhibited substantial absolute deviations of 0.95±0.19 Gy[RBE] and 1.20±0.27 Gy[RBE] in Experiment 1, which were reduced to 0.67±0.13 Gy[RBE] and 0.94±0.28 Gy[RBE] in Experiment 2 and further decreased to 0.45±0.18 Gy[RBE] and 0.68±0.14 Gy[RBE] in Experiment 3, respectively. The improvement is statistically significant for most reference DVH indices with $p < 0.05$ using noisy probing dose method as shown in Figure 3(a).

Additionally, the absolute deviations of Dmean of spaceOAR — an implant utilized in prostate cancer patients treated with PBSPT for rectal protection — were measured to be 0.51±0.22 Gy[RBE], 0.36±0.12 Gy[RBE], and 0.19±0.08 Gy[RBE] for the three experiments respectively. This signifies that even in the presence of the implant (a non-human-being issue), the dose prediction accuracy remained commendable, with further improvements observed in Experiment 3.

3.3 3D Gamma Passing Rates Evaluation

Figure 3(b) displays the 3D Gamma passing rates (3%/2mm/10%) for targets, OARs, and the BODY (regions outside the ROI) for prostate and lung test cases, respectively. Compared to Experiment 1, Gamma passing rates exhibited improvements for both targets and OARs in Experiment 2. The Gamma passing rates (mean ± standard deviation) for prostate targets escalated from 95.03%±1.18% to 97.95%±1.15%, while the improvement for lung targets was from 91.55%±1.27% to 95.34%±1.17%. In terms of prostate OARs, the increase was from 91.15%±1.53% to 95.01%±1.21%, and for lung OARs, the Gamma pass rate rose from 88.78%%±1.83% to 93.02%±1.66%. Furthermore, the improvement in Gamma passing rates for the BODY region in Experiment 2 was even more pronounced compared to targets and OARs. For the prostate BODY, the Gamma passing rate elevated from 84.81%±1.36% to 92.75%±1.40%, and for the lung BODY, the rate advanced from 83.79%±1.01% to 90.93%±1.29%. In Experiment 3, the Gamma pass rates witnessed an additional increase of approximately 1.7% for targets and an average of 2.9% for OARs and BODY relative to the 3D Gamma passing rates from Experiment 2. The 3D Gamma passing rate improvement is statistically significant for target, OARs and BODY with p<0.05 using noisy probing dose method as shown in Figure 3(b).

## 3.4 Dice Coefficient Evaluation

Figure 4 illustrates the dice coefficients for volumes encompassed by the iso-dose lines (with dose values varying from 10% to 90% of the prescribed dose at intervals of 10%) between the predicted and ground truth dose distributions for the prostate and lung test cases. In general, the average dice coefficients demonstrated an improvement for the iso-dose lines in Experiment 2 in comparison to Experiment 1. Further improvements were observed in Experiment 3. With respect to high percentage iso-dose lines, modest improvements were observed across the three experiments. For example, the 90% iso-dose lines noted values of 0.953±0.007, 0.971±0.003, and 0.983±0.005 for prostate, and 0.931±0.021, 0.947±0.019, and 0.967±0.015 for lung across the three experiments,

respectively. For lower percentage iso-dose lines, more significant improvements were observed among the three experiments. For instance, the 10% iso-dose lines reported values of 0.893±0.019, 0.931±0.015, and 0.957±0.019 for prostate, and 0.870±0.023, 0.917±0.019, and 0.945±0.021 for lung across the three experiments, correspondingly.

3.5 Uncommon Cases Test

Figure 5 presents the 3D Gamma passing rates within targets, OARs, and BODY for the uncommon test cases with different beam settings and/or dose prescriptions from the training datasets. Compared to Experiment 1, Experiment 2 demonstrated improved Gamma passing rates for both targets and OARs. The Gamma passing rates for prostate targets increased from 89.32%±1.45% to 93.48%±1.51%, and for lung targets, the Gamma passing rates improved from 85.93%±2.01% to 90.08%±1.34%. For prostate OARs, the improvement was from 85.87%±1.73% to 91.15%±1.13%, while for lung OARs, the Gamma passing rates climbed from 83.42%±1.76% to 89.45% ± 1.32%. Notably, the improvement in Gamma passing rates for the BODY region in Experiment 2 was more significant compared to targets and OARs. Specifically, for the prostate BODY, the Gamma passing rate ascended from 79.19%±2.03% to 88.71%±1.12%, and for the lung BODY, the Gamma passing rate grew from 77.49%±1.05% to 86.53%±1.32%. In Experiment 3, the Gamma pass rates demonstrated further improvement of approximately 3.3% for targets and an average boost of 4.5% for OARs and BODY, compared to the rates from Experiment 2. The 3D Gamma passing rate improvement is statistically significant for all uncommon cases with $p < 0.05$ using noisy probing dose method as shown in Figure 5.

a)

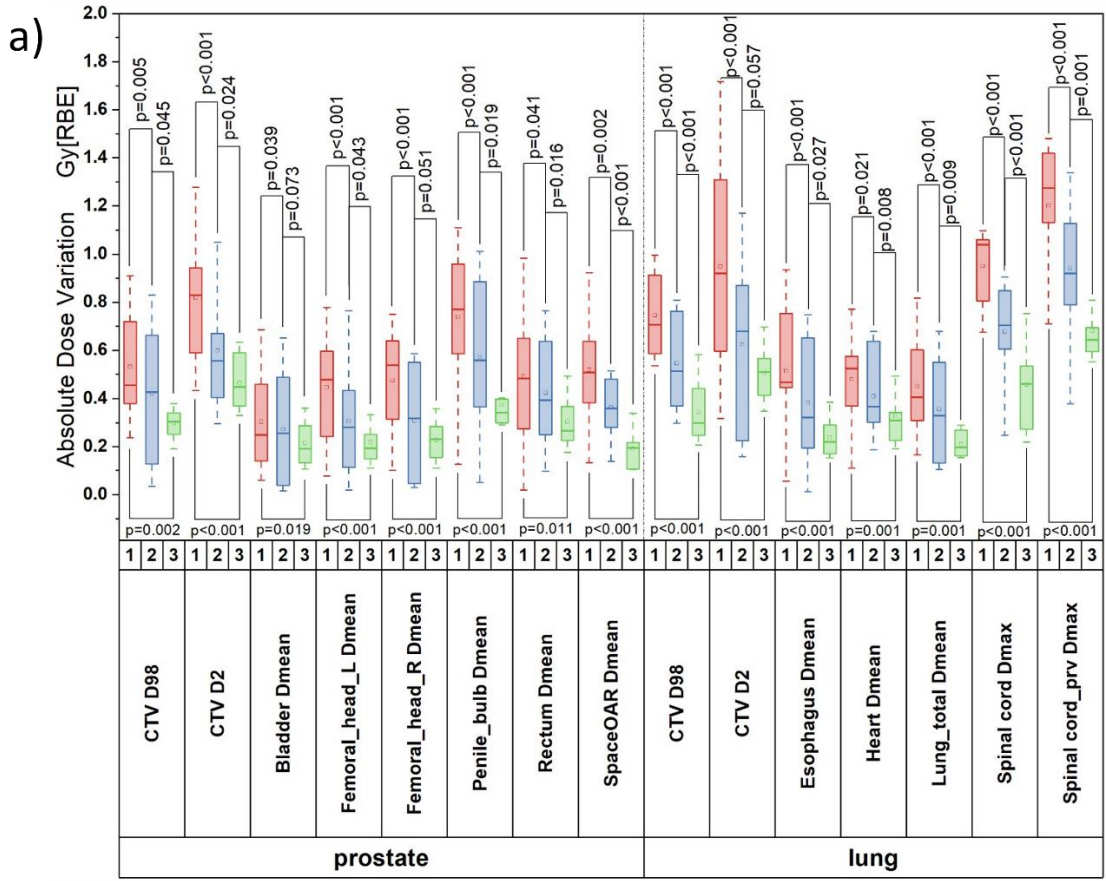

b)

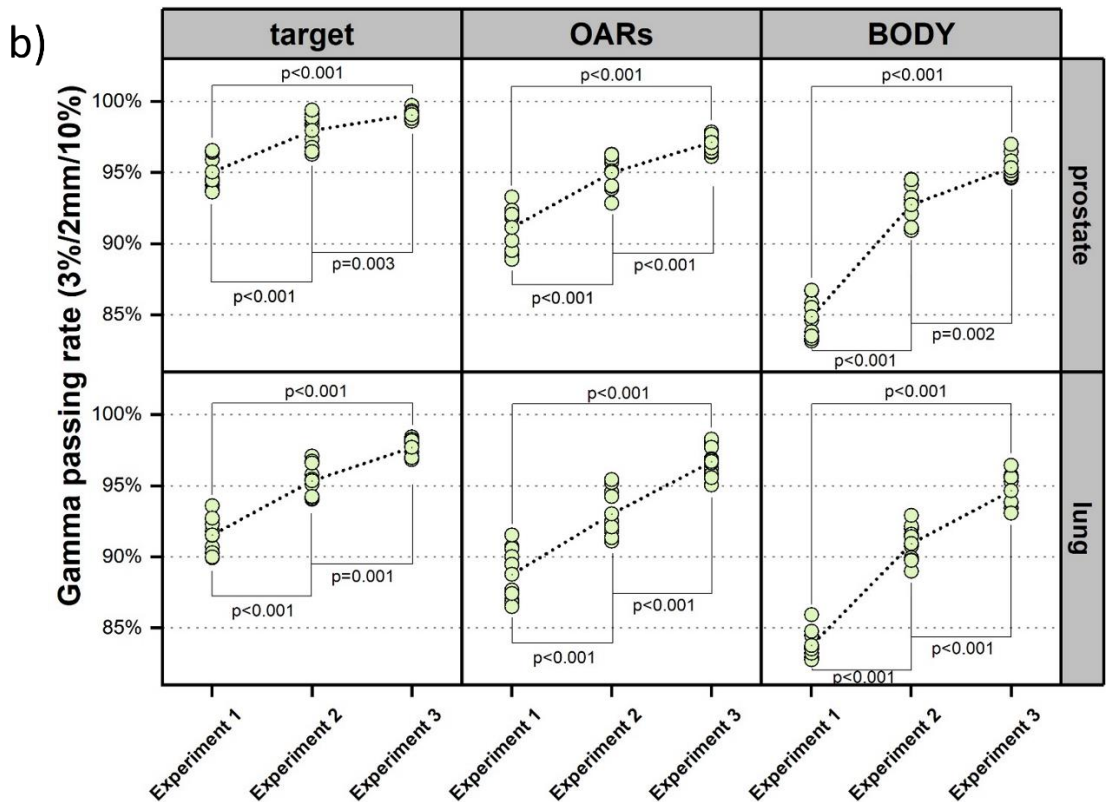

Figure 3. a) Boxplot (depicting the minimum, first quartile, median, third quartile, and maximum values) showcasing the absolute divergence of the DVH indices for targets and OARs between the ground truth and predicted doses from three distinct experiments for prostate and lung test cases. In the diagram, Experiment 1, Experiment 2, and Experiment 3 are represented by colors red, blue, and green, respectively. b) 3D Gamma passing rates (3%/2mm/10%) within targets, OARs, and the BODY (targets and OARs excluded) for prostate and lung test cases, with 10 cases from each testing group represented by green circles. The dotted lines in the figure symbolize the trend of the average value for all the test cases within each respective group. P-values, derived from the Wilcoxon signed rank tests, were also presented for comparisons between Experiment 1 and 2, Experiment 2 and 3, as well as Experiment 1 and 3.

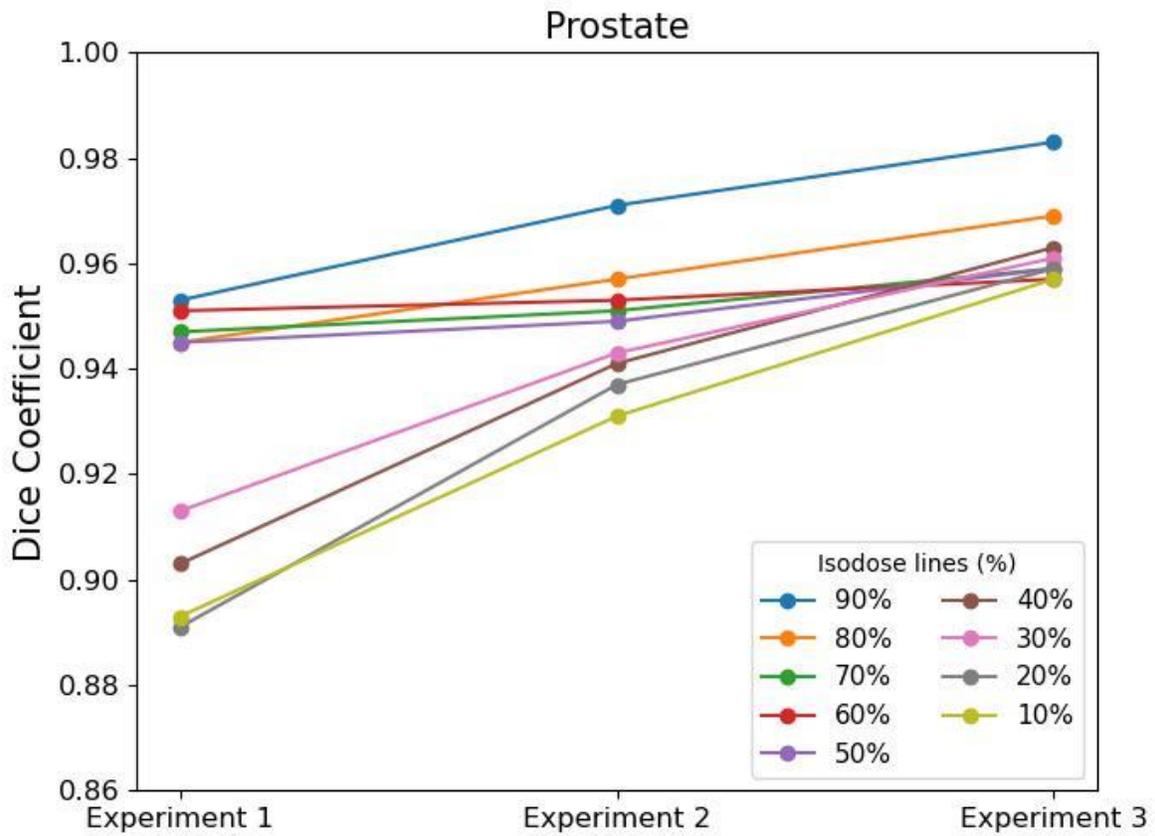

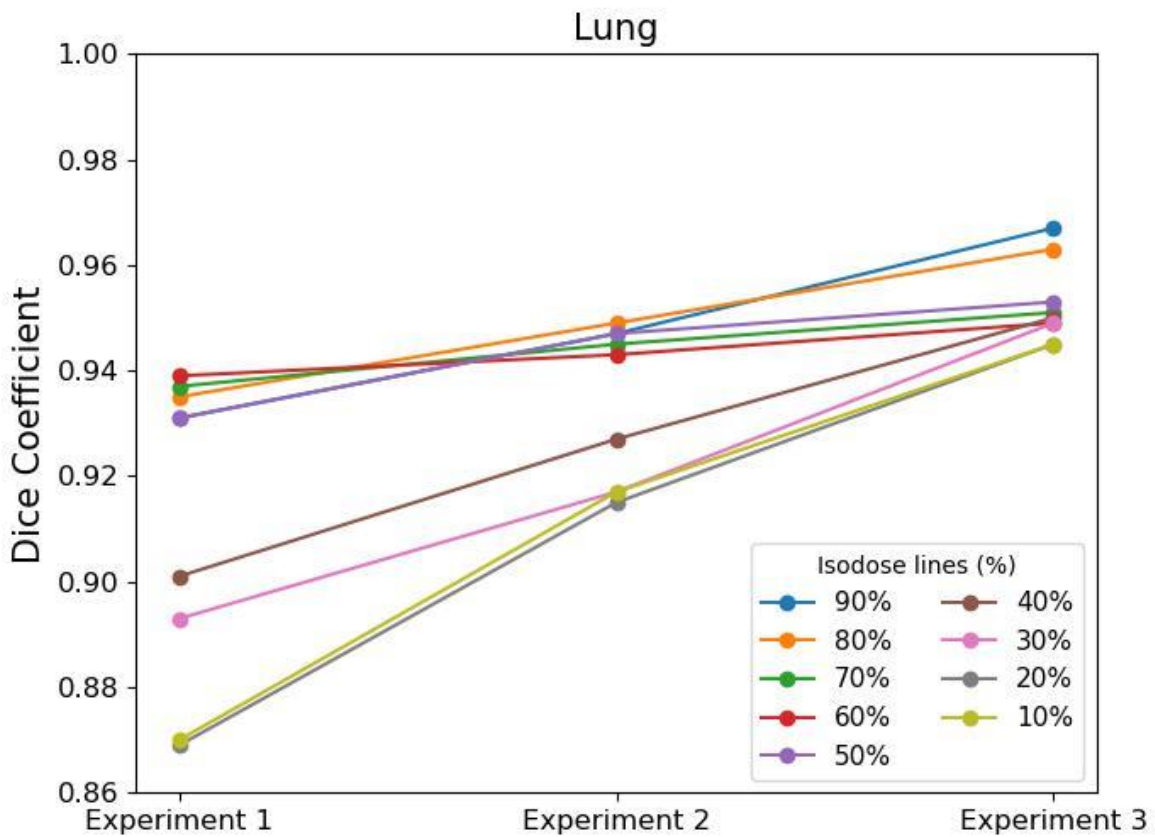

Figure 4. Dice coefficients for structures confined by the iso-dose lines, which range from 10% to 90% of the prescribed dose in increments of 10%. The coefficients are derived from comparisons between the predicted and ground truth doses for prostate (Fig. 4(a)) and lung (Fig. 4(b)) test cases.

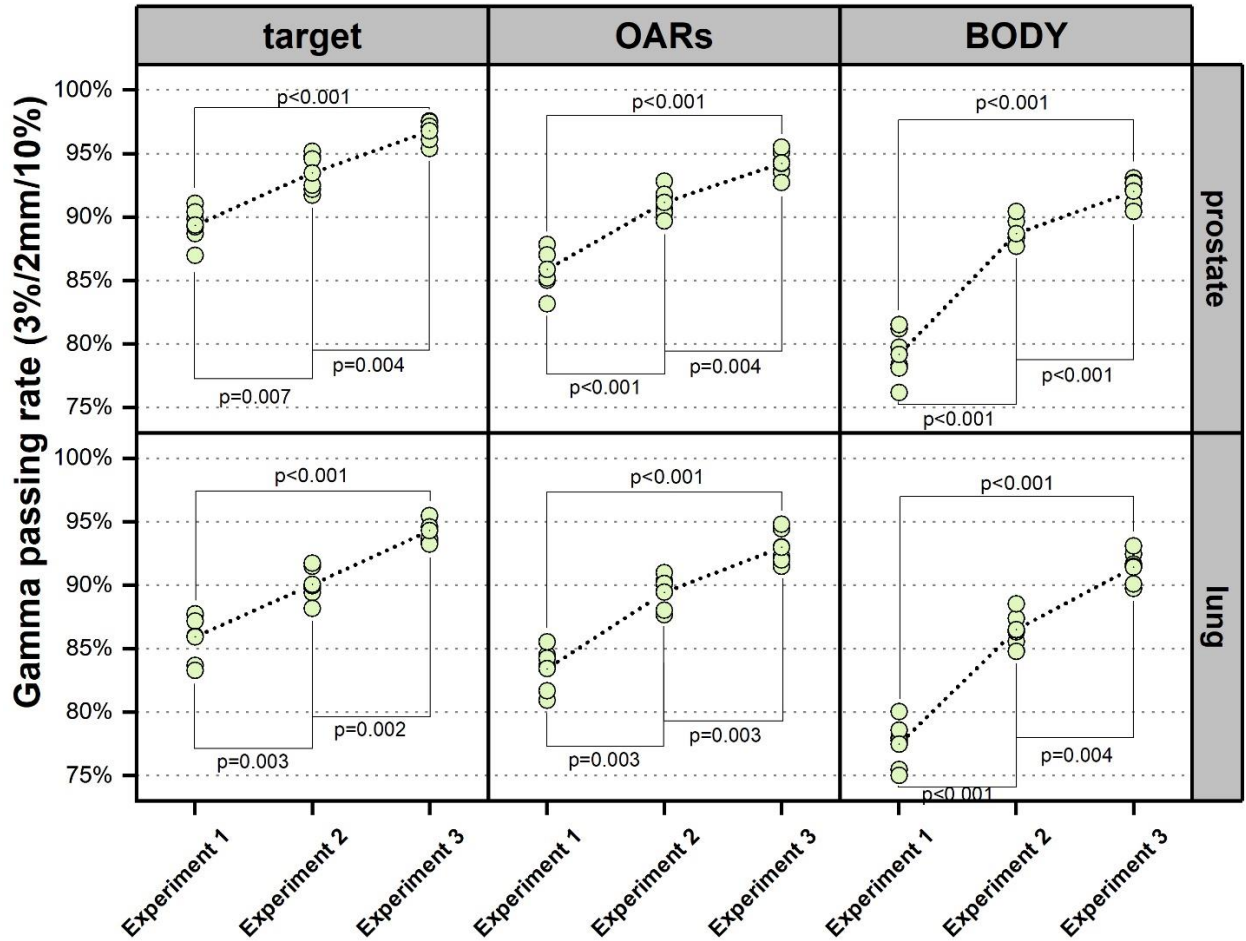

Figure 5. 3D Gamma passing rates (3%/2mm/10%) within targets, OARs, and the BODY (outside ROI) for prostate and lung uncommon test cases, with 6 cases from each site represented by green circles. The dotted lines in the figure symbolize the trend of the average value for all the test cases within each respective group. P-values, derived from the Wilcoxon signed rank tests, were also presented for comparisons between Experiment 1 and 2, Experiment 2 and 3, as well as Experiment 1 and 3.

## 4. Discussion

In this study, we investigated the utilization of the noisy probing dose method for dose prediction in PBSPT by comparing it with the conventional ROI-based method and the recently proposed beam mask and sliding window method. The noisy probing dose method demonstrated notable improvements in DVH index accuracy, including both D2, D98 for targets and Dmean and Dmax for OARs.

We evaluated the spatial agreement between the predicted and ground truth doses using 3D Gamma analysis and dice coefficients. While many studies prioritize dose agreement within the ROI only, we emphasized accuracy in targets, OARs, and the BODY (outside ROI). Our novel noisy probing dose method surpassed the previously reported methods, especially outside the ROI. The consistent isodose volumes enclosed by different percentages of the prescription dose highlight its superior dose prediction both inside and outside the ROI, demonstrating a holistic grasp of the entire dose gradient.

AI models, including dose prediction, can fail or underperform clinically due to outlier cases not represented in their training dataset, like patients with uncommon beam configurations or dose prescriptions. This issue has not been fully addressed in the previous literature[33-36,38,39,44-47]. In this study we used six outlier cases each with uncommon beam setups and/or prescription doses not existing in the model training dataset, to further test the performance of the proposed noisy probing dose model. From the results, it can be clearly seen that when dealing with outlier cases, the Gamma passing rates obtained by the conventional ROI-based method are much lower than the clinically acceptable standard. Although the beam mask and sliding window method showed improvement, they still did not meet the clinically acceptable standard. On the other hand, the Gamma passing rates obtained by the proposed noisy probing dose model approached the clinically acceptable standard of a Gamma passing rate of 95% in both target and OAR regions, reflecting the model's robustness in handling uncommon clinical scenarios[50]. This can be explained by the fact that the noisy probing dose model carries more physics information, allowing the model to learn more knowledge about the physics interaction between protons and medium in proton dose calculation, such as dose fall-off. This implies that the model can be more generalizable than the previous methods, exhibiting robust performance even when handling outlier clinical cases.

For a long time, AI-based dose prediction methods, especially those directly predicting doses based on the ROI, have been considered as non-physics methods, raising concerns about their generalizability[29]. This leads to two concerns: the accuracy of the dose prediction results for high-precision irradiation techniques such as PBSPT, and their ability to generalize dose predictions, i.e., the extent to which the model can understand and predict dose distributions accurately for some outlier clinical cases. Current research attempts to enhance the model's accuracy and robustness by including different parameters like beam angle[44-47]. However, considering the multitude of physics parameters influencing the final dose, it's impractical to include all of them in the model training. Moreover, adding more parameters into the model training significantly increases the model complexity, making it challenging for effective clinical deployment and resulting in possible overfitting. Our proposed noisy probing dose method took inspiration from the conventional influence matrix concept used in dose calculation and optimization. We used a noisy probing dose to represent all the physics information required by the AI-based methods, such as beam angles and dose fall-off, etc. As Figure 1 shows, we've established an AI-based dose prediction workflow centered around the noisy probing dose concept.

Our results have demonstrated that the proposed noisy probing dose method can make accurate dose predictions in prostate and lung cancer patients treated with PBSPT. More importantly, the noisy probing dose method can make clinically acceptable and accurate dose predictions for uncommon clinical cases, showcasing its excellent generalizability capability stemming from the underlining physics. The results also indicate the vital role of incorporating physics to train an AI model for accurate dose predictions, particularly for improving the model's generalizability to handle outlier cases. Additionally, compared to these two findings, our model is notable for its technical feasibility, balancing precision with ease of implementation. Drawing upon the concept and calculations of the influence matrix from traditional radiotherapy treatment planning, we employed a noisy non-modulated dose as the probing dose. This approach fully harnesses the analytical prowess of AI. While this method imbues the model with robust inherent physics logic, it also simplifies its practical application in routine clinical settings.

## 5. Conclusion

In conclusion, we have developed a highly accurate PBSPT dose prediction method and have tested it in prostate and lung cancer patients based on a novel concept of noisy probing dose. This method can not only make accurate dose prediction both within and outside the ROI, but it also provides clinically acceptable dose prediction for outlier clinical cases. With more physics information included, the proposed noisy probing dose method enhances the accuracy of dose prediction for PBSPT, endowing them with excellent generalizability. It also strikes a balance between maximal accuracy and limited complexity, making it ready to be implemented clinically.